\documentclass[english]{elsarticle}

\usepackage{times}
\usepackage[T1]{fontenc}
\usepackage[latin1]{inputenc}
\usepackage{amsmath}
\usepackage{amssymb}	
\usepackage{MnSymbol}
\usepackage{xfrac}
\usepackage{float}
\usepackage{color}
\usepackage{graphicx}
\usepackage{xfrac}
\makeatletter
\usepackage{babel}
\makeatother

\begin{document}

\title{Cooper Pair Insulator phase induced in amorphous Pb$_\mathbf{0.9}$Bi$_\mathbf{0.1}$ thin films}

\date{\today{}}
\author[bphys]{{S. M. Hollen}\corref{cor1}\fnref{fn1}}\ead{hollen.8@osu.edu}
\author[bphys]{{J. Shainline}\fnref{fn2}}
\author[bphys,bengin]{J. M. Xu} 
\author[bphys]{J. M. Valles, Jr.}

\address[bphys]{Department of Physics, Brown University, Providence, RI 02912}
\address[bengin]{School of Engineering, Brown University, Providence, RI 02912}

 \cortext[cor1]{Corresponding author}
 \fntext[fn1]{Present address: Center for Emergent Materials, The Ohio State University, Columbus, OH, 43210.}
 \fntext[fn2]{Present address: Department of Electrical, Computer, and Energy Engineering, University of Colorado Boulder, Boulder, CO, 80309.}
 
\begin{abstract}
A Cooper pair insulator (CPI) phase emerges near the superconductor-insulator transitions of a number of strongly-disordered thin film systems. Much recent study has focused on a mechanism driving the underlying Cooper pair localization.  We present data showing that a CPI phase develops in amorphous Pb$_{0.9}$Bi$_{0.1}$ films deposited onto nano-porous anodized aluminum oxide surfaces just as it has been shown to develop for a-Bi films.  This result confirms the assertion that the CPI phase emerges due to the structure of the substrate. It supports the picture that nanoscale film thickness variations induced by the substrate drive the localization.  Moreover, it implies that the CPI phase can be induced in any superconducting material that can be deposited onto this surface.
 
\end{abstract}

\begin{keyword} 
Cooper pair insulator, 
superconducting islands, 
Little-Parks oscillations,
superconductor-insulator transition, 
nanoscale patterning, 
thin film.
\end{keyword}

\maketitle

An insulating phase exhibiting a number of curious transport properties that can be attributed to the presence of localized Cooper pairs (CPs) appears in several highly-disordered thin film systems near their superconductor-insulator transitions (SITs)\cite{Gantmakher:JETP1998, Sambandamurthy:PRL2004, Steiner:PRL2005, Baturina:PRL2007, Sacepe:NatPhys2011, Stewart:Science2007, Mondal:PRB2012}. The hallmark of this CP insulator (CPI) phase is a giant magnetoresistance (MR) peak that appears at temperatures below 1K in perpendicular magnetic fields.   CP insulators also exhibit thermally activated transport at the lowest temperatures and reentrant dips in their $R(T)$.  The qualitative resemblance of these behaviors to those exhibited by insulating arrays of weakly coupled superconducting islands \cite{Delsing:PRB1994,Fistul:PRL2008,Syzranov:PRL2009} has led to the general belief that these phenomena result from transport by incoherent Cooper Pair tunneling between localized states that becomes more incoherent in magnetic field. Magnetotransport studies of amorphous Bi (a-Bi) films patterned with a nanohoneycomb (NHC) array of holes tended to confirm this picture\cite{Stewart:Science2007, Nguyen:PRL2009}.  Those experiments showed similar MR peaks and thermally activated resistances.  Most significantly, they revealed that the thermal activation energy oscillates with a period dictated by the Cooper pair or superconducting flux quantum\cite{Stewart:Science2007,Welp:PRB2002}.  Thus, multiple theoretical models presume that weakly coupled islands of Cooper Pairs exist in the CPI phase\cite{Dubi:Nature2007, Feigelman:APhys2010, Falco:PRL2010}.  Results of calculations based on these models are consistent with the presence of CP islands in the films and show similarities to models of weakly-disordered Josephson junction arrays\cite{Syzranov:PRL2009}. Also, recent scanning tunneling spectroscopy experiments in TiN\cite{Baturina:PRL2007} and In oxide (InO$_{x}$)\cite{Gantmakher:JETP1998, Sambandamurthy:PRL2004, Steiner:PRL2005} films revealed nanoscale superconducting inhomogeneities that are likely precursors to island formation\cite{Bouadim:NatPhys2011}.  The origin of these islands, however, has not been established\cite{Gantmakher:PUsp2010}.
 
With the establishment of the CPI phase, more attention focuses on how the localized Cooper pair islands form.  Quantum Monte Carlo calculations suggest that very strong uniform electronic disorder induces their appearance\cite{Ghosal:PRL1998, Bouadim:NatPhys2011}.  
Recently, nanoscale film thickness inhomogeneities have been argued to induce the CPI phase observed in NHC a-Bi films\cite{Hollen:PRB2011,Hollen:JPhys2012}. 
This latter work provides a model for how the CPI state forms in NHC films that relies on the substrate topography and not on the elements in the evaporated film.  We tested this model by looking for CPI behavior in another film system grown in the same way, a-Pb$_{0.9}$Bi$_{0.1}$ (a-PbBi) NHC films.

The driver of CP localization in NHC films appears to be spatial variations in their thickness.  These variations are inferred from analysis of AFM topographs of the anodized aluminum oxide (AAO) substrates used as templates for the NHC films. The topographs revealed regular, nanoscale height variations\cite{Hollen:PRB2011}.  Since the local deposition thickness depends on the local slope of the substrate, these height variations lead to film thickness variations.  Analysis of data from multiple substrates indicates that close to the thickness-driven SIT, the thickest film regions form islands that can support Cooper pairing. At the SIT, these islands become linked by regions that are just thick enough for pairing.  This model predicts that materials other than Bi exhibit a CPI phase provided they grow similarly. Here we present a test of that prediction using PbBi.  

We present data that show that this new a-PbBi NHC film system exhibits all the characteristics of the CPI state, including simply activated insulators, reentrance in films near the thickness-driven SIT (d-SIT), and a giant MR peak in films near the SIT.  Additionally, we report the temperature and disorder dependence of $H_\mathrm{peak}$ and $R_\mathrm{peak}$, and the field dependence of the activation energy, $T_{0}$, and prefactor, $R_{0}$, for these films. The results imply that a CPI phase can be induced in any elemental superconductor by this method. Consequently, previously fixed properties of Cooper pair insulators, such as the pairing strength, can now be varied by choosing different materials.

Amorphous PbBi NHC films were grown using the same methods as for previously studied a-Bi NHC films.  PbBi was thermally evaporated onto a substrate of anodized aluminum oxide (AAO) held at 8K on the mixing chamber stage of a dilution refrigerator.  The substrate was pre-coated with a 10nm Ge underlayer and Au/Ge contact pads at room temperature.  A 1nm layer of Sb was quench condensed onto the prepared substrate before subsequent layers of PbBi to ensure amorphous film growth\cite{Ekinci:PRB1998}.  Film thicknesses were measured with a calibrated quartz crystal monitor. The resulting films acquire the NHC geometry of the underlying AAO substrate, and also develop thickness variations due to the changing local slope of the AAO (see ref. \cite{Hollen:PRB2011}). The holes form a roughly triangular array with an average hole radius of 17$\pm$9nm and spacing of 85$\pm$30nm (see Fig. 1 inset).  Film sheet resistances, $R_{\square}$, were measured using standard four-point AC and DC techniques in the linear response regime on a (1mm)$^{2}$ area of film. 
  
The transport characteristics of a-PbBi NHC films shown in Figs. 1 and 2 are very similar to those of a-Bi NHC films\cite{Stewart:Science2007,Nguyen:PRL2009}. Fig. 1a shows a series of films traversing the thickness-driven SIT on an Arrhenius scale, where the thinner films are insulators (defined by $dR/dT<0$ at the lowest $T$) and the thickest film is a superconductor. Insulating films are simply activated at low temperatures with $R=R_{0}e^{T_{0}/T}$. The activation energy, $T_{0}$, decreases linearly to zero with increasing thickness (Fig. 1c).  The extrapolation to $T_{0}=0$ marks the critical thickness for the SIT.  The critical normal state ($T=$8K) sheet resistance $R_{N,c}$ is close to 18k$\Omega$.  Finally, the linearly increasing normal state conductance $G_{N}=1/R_{N}$ with film thickness (Fig. 1b) verifies that the film growth is amorphous\cite{Liu:PRB1993,Dynes:PRL1978}.

The reentrance feature that develops in the $R(T)$ near the d-SIT appears more clearly on a linear temperature scale, as in Fig. 2a. This feature implies the presence of localized Cooper pairs in the insulating films, which is confirmed by the observation of Little-Parks-like MR oscillations at low fields (Fig. 2b, inset)\cite{Stewart:Science2007}.  With an incremental increase in thickness, the last reentrant film gives way to a superconducting film with a very broad transition region and a remarkably high $T_{c}$ ($\approx$ 1K).    

Most importantly for the identification of the CPI phase, the films near the SIT exhibit the giant MR peak that has become its signature (Fig. 2b).  Even at the relatively high temperature of 300mK, the resistance of film 5 grows to more than half a megaohm at $H_\mathrm{peak}\simeq 2.6$T, which is more than 5 times its zero-field value.  The position of the MR peak, $H_\mathrm{peak}$, shifts to lower fields with decreasing thickness (Figs. 2b), as also previously observed in a-Bi NHC films\cite{Nguyen:PRL2009}.  The insulators become more strongly activated in field up to $H=H_\mathrm{peak}$ and then weaken beyond $H_\mathrm{peak}$ (Fig. 3a).  Because of this temperature dependence, the giant MR peak and oscillations grow dramatically with decreasing temperature, as shown in Fig. 3d.  These features are mirrored in the field-dependence of the activation energy, $T_{0}(H)$, while the prefactor, $R_{0}$, increases monotonically with field. (Fig. 3b and c).  At the highest fields, fits to activated behavior begin to fail as weakly localized insulating behavior, $1/R(T)=G(T)\propto \ln(T)$, emerges, as shown in Fig. 4. These observations are all consistent with those for the CPI phase previously reported for a-Bi NHC films. 

The appearance of this CPI phase in a-PbBi films can be attributed to the properties of the AAO substrate.  Like a-Bi films, a-PbBi films on nominally flat, unpatterned fire-polished glass subtrates do not show signs of a localized Cooper pair phase.  Thus, the present results strongly suggest that the Cooper pair insulator phase can be induced in other amorphous film systems by depositing them on AAO.  Moreover, they provide additional support for the model that the nanoscale height variations on the surface of the AAO substrate induce film thickness variations that drive the Cooper pair localization. 

We gladly thank A. Berg and J. C. Joy for their assistance with the data collection.  This work was supported by the NSF through Grants No. DMR-0605797 and No. DMR-0907357, by the AOARD, the AFOSR, and the WCU program at SNU, Korea. We are also grateful for the support of AAUW.

\begin{figure}
\begin{center}
\includegraphics[width=.5\columnwidth,keepaspectratio]{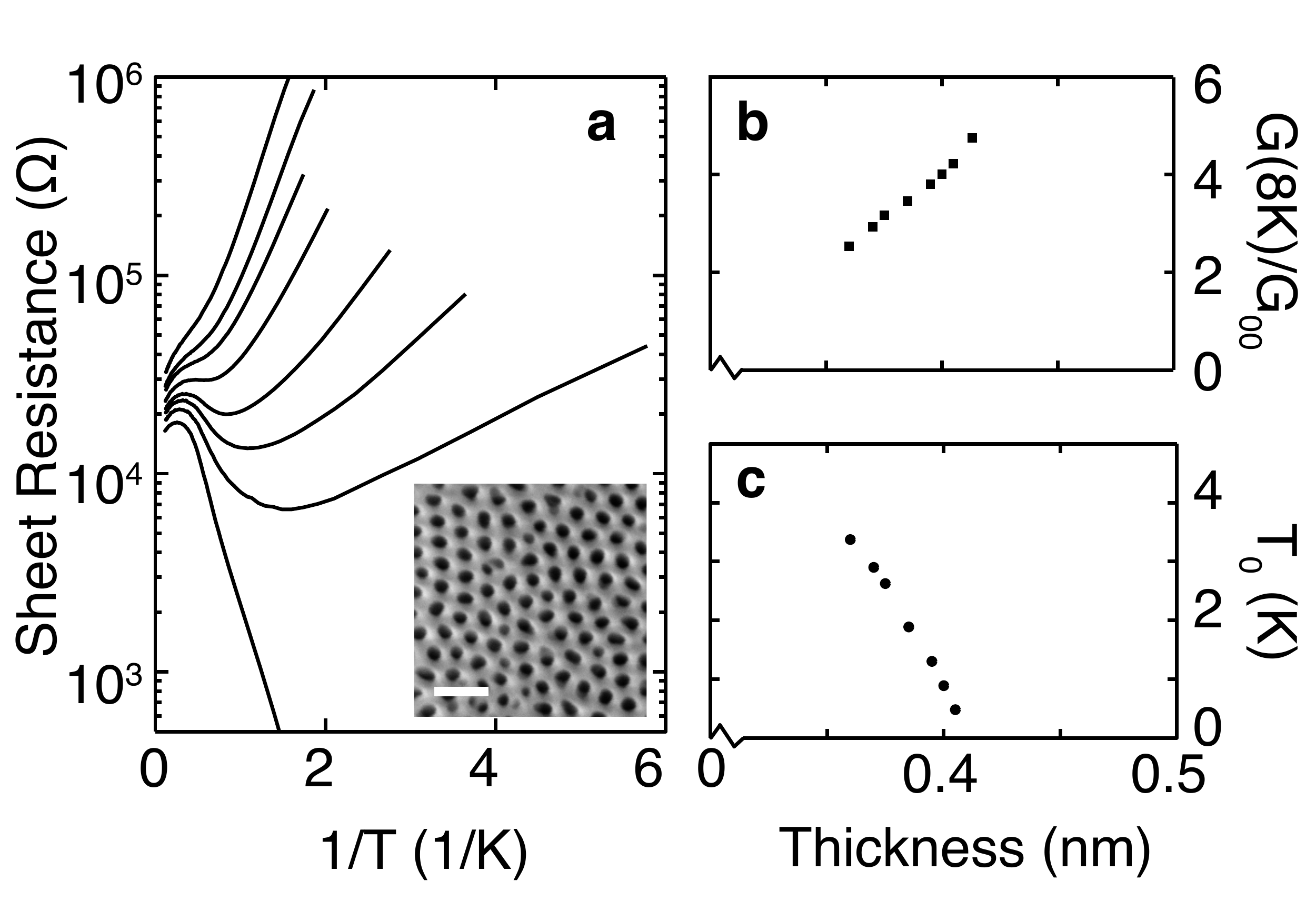}
\caption{a)  Sheet resistance dependence on temperature shown on an Arrhenius scale for a series of amorphous Pb$_{0.9}$Bi$_{0.1}$ films deposited on the NHC substrate shown in the inset (scale bar is 200nm) Film thicknesses from top to bottom are 0.36, 0.37, 0.375, 0.385, 0.395, 0.4, 0.405, and 0.413 $\pm$ 0.01 nm.  b)  The normal state sheet conductance, $G_{N\square}(8K)=1/R_{N\square}(8K)$, normalized by the quantum of conductance $G_{00}=81k\Omega$ is linear with thickness, indicating amorphous film growth.  c) The activation energy, $T_{0}$, obtained by fits to $R(T)\propto R_{0}e^{T_{0}/T}$ in (a), decreases nearly linearly with thickness, crossing $T_{0}=0$ at the SIT.
\label{cap:fig1}}
\end{center}
\end{figure}

\begin{figure}
\begin{center}
\includegraphics[width=1\columnwidth,keepaspectratio]{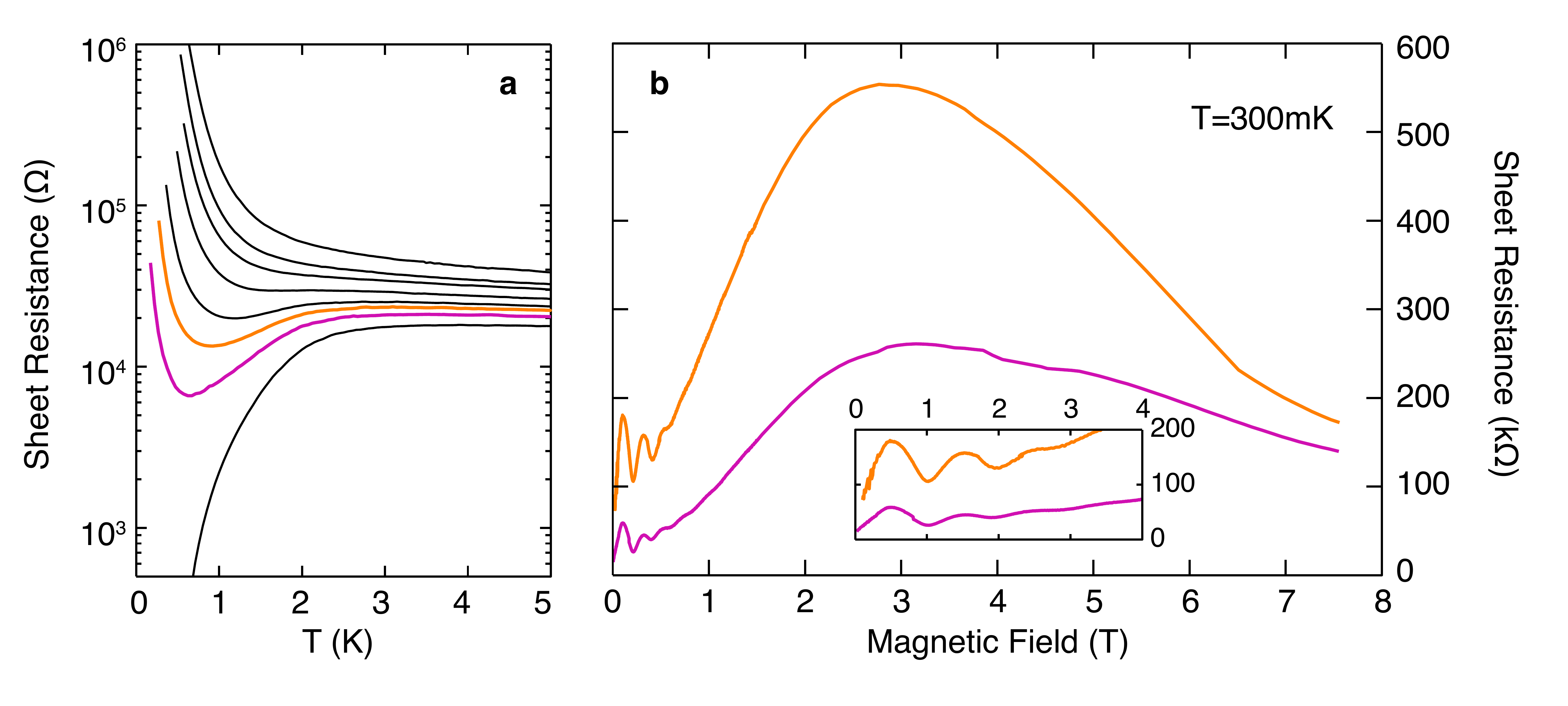}
\caption{a) Sheet resistance dependence on temperature for the same series of films as Fig. 1a, but shown here on a linear temperature scale.  b) Magnetoresistance for a field applied perpendicularly to the plane of the film at 300mK for films 5 and 6 of (a).  Inset: a magnified view of the low-field region showing the magnetoresistance oscillations.  The x-axis is labeled in units $f=H/H_\mathrm{M}$ where $H_\mathrm{M}$ corresponds to a magnetic flux density of one flux quantum per hole. 
\label{cap:fig2}}
\end{center}
\end{figure}

\begin{figure}
\begin{center}
\includegraphics[width=.5\columnwidth,keepaspectratio]{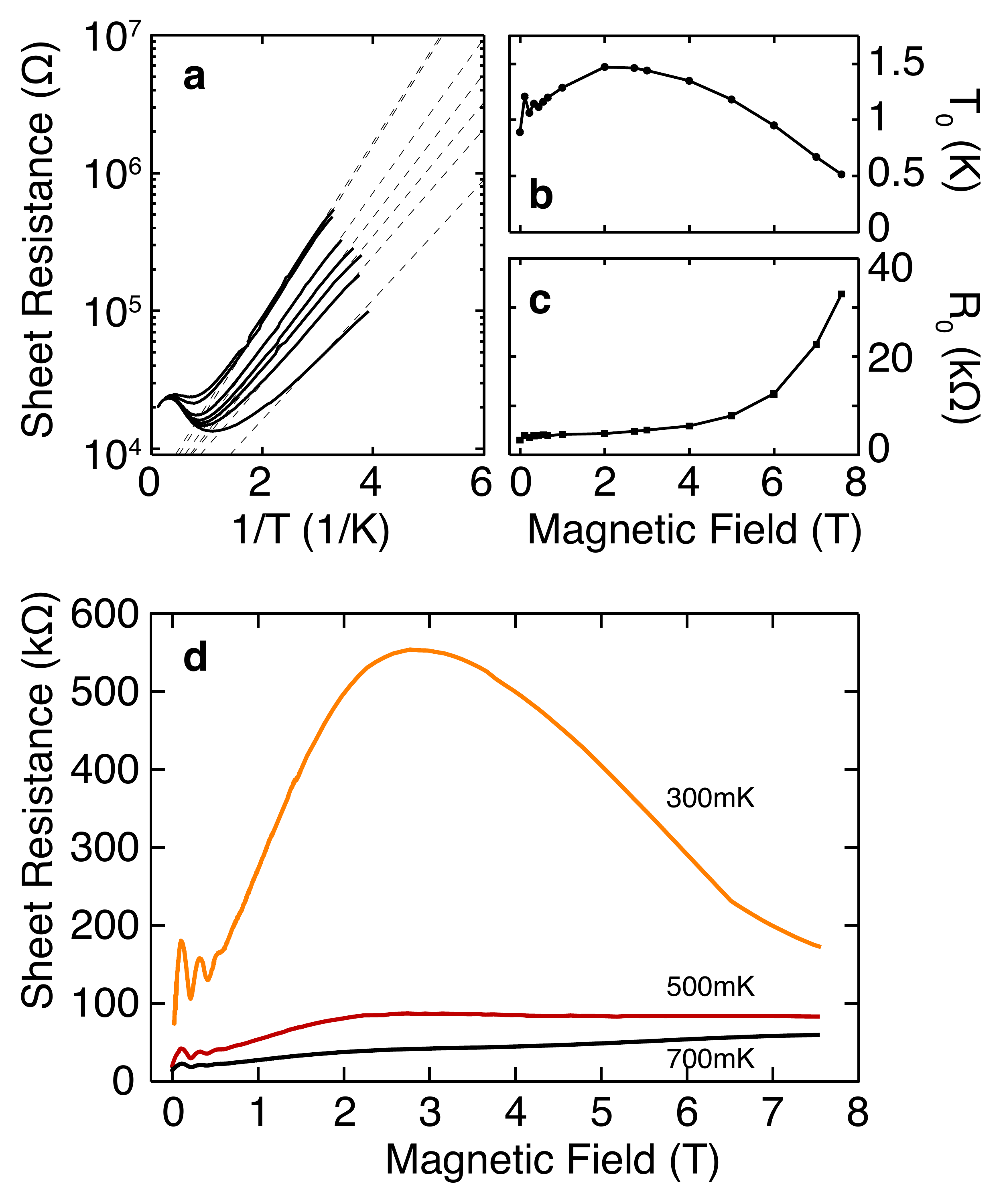}
\caption{a) Sheet resistance dependence on temperature film 5 at perpendicularly applied magnetic fields of 0, 0.22 ($H_\mathrm{M}$), 0.44 ($2H_\mathrm{M}$), 0.66 ($3H_\mathrm{M}$), 1.00, 2.00, 2.70T ($\simeq H_\mathrm{peak}$) (from bottom to top). Dashed lines are fits to activated behavior.  b) Activation energy $T_{0}$ and c) prefactor $R_{0}$ dependence on magnetic field. d) Magnetoresistance of film 5 of Fig. 1a (and Fig. 2a) at 300, 500 and 700mK.  
\label{cap:fig3}}
\end{center}
\end{figure}

\begin{figure}
\begin{center}
\includegraphics[width=.5\columnwidth,keepaspectratio]{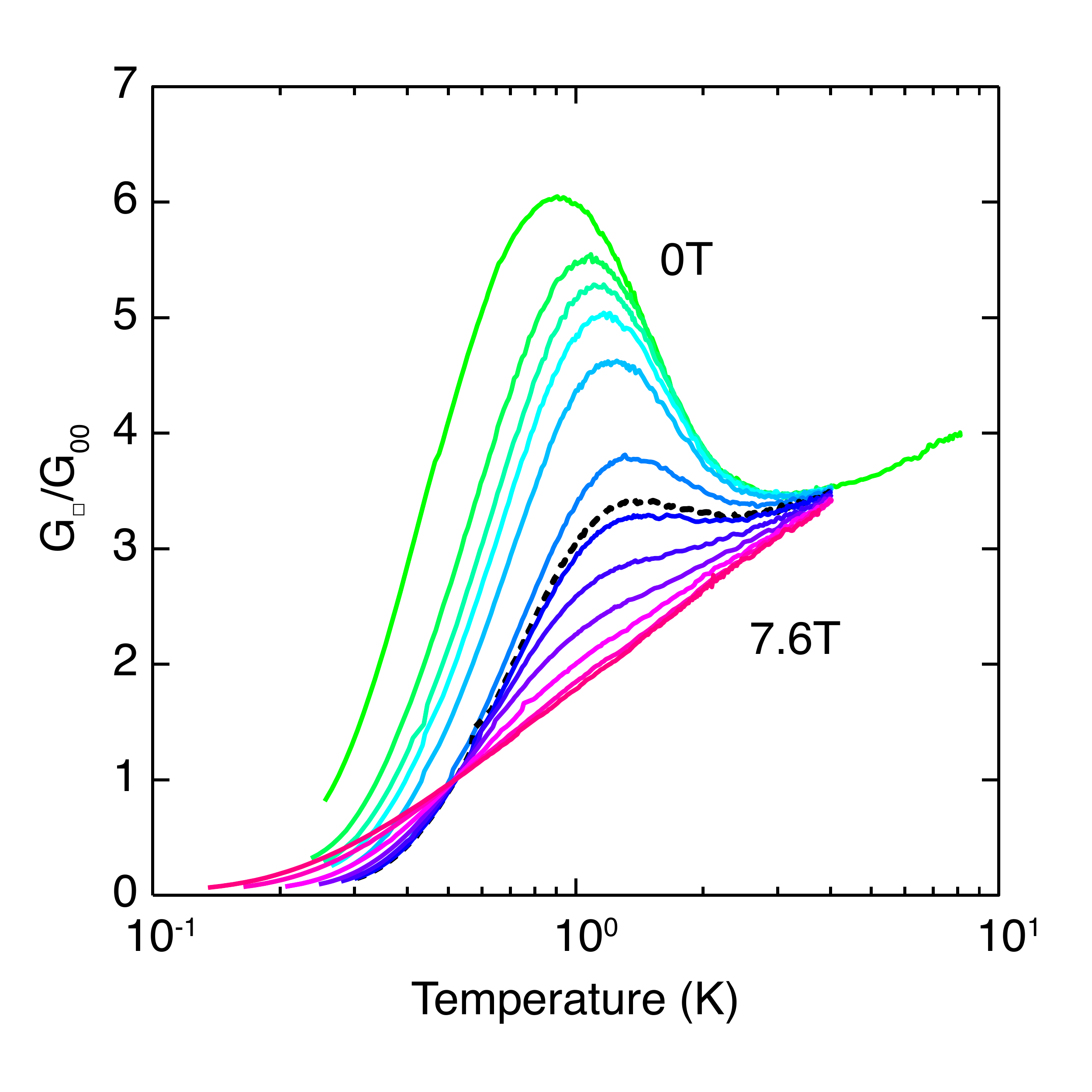}
\caption{Normalized sheet conductance ($G_{00}=81k\Omega$)  of film 5 versus temperature for a series of perpendicularly applied magnetic fields with values: 0, 0.22 ($H_\mathrm{M}$), 0.44 ($2H_\mathrm{M}$), 0.66 ($3H_\mathrm{M}$), 1, 2, 2.7 ($\simeq H_\mathrm{peak}$, dashed black line), 3, 4, 5, 6, 7, and 7.6T. 
\label{cap:fig5}}
\end{center}
\end{figure}


\begin{thebibliography}{23}
\expandafter\ifx\csname natexlab\endcsname\relax\def\natexlab#1{#1}\fi
\expandafter\ifx\csname bibnamefont\endcsname\relax
  \def\bibnamefont#1{#1}\fi
\expandafter\ifx\csname bibfnamefont\endcsname\relax
  \def\bibfnamefont#1{#1}\fi
\expandafter\ifx\csname citenamefont\endcsname\relax
  \def\citenamefont#1{#1}\fi
\expandafter\ifx\csname url\endcsname\relax
  \def\url#1{\texttt{#1}}\fi
\expandafter\ifx\csname urlprefix\endcsname\relax\def\urlprefix{URL }\fi
\providecommand{\bibinfo}[2]{#2}
\providecommand{\eprint}[2][]{\url{#2}}

\bibitem[{\citenamefont{Gantmakher et~al.}(1998)\citenamefont{Gantmakher,
  Golubkov, Dolgopolov, Tsydynzhapov, and Shashkin}}]{Gantmakher:JETP1998}
\bibinfo{author}{\bibfnamefont{V.~F.} \bibnamefont{Gantmakher}},
  \bibinfo{author}{\bibfnamefont{M.~V.} \bibnamefont{Golubkov}},
  \bibinfo{author}{\bibfnamefont{V.~T.} \bibnamefont{Dolgopolov}},
  \bibinfo{author}{\bibfnamefont{G.~E.} \bibnamefont{Tsydynzhapov}},
  \bibnamefont{and} \bibinfo{author}{\bibfnamefont{A.~A.}
  \bibnamefont{Shashkin}}, \bibinfo{journal}{JETP Lett.}
  \textbf{\bibinfo{volume}{68}}, \bibinfo{pages}{363} (\bibinfo{year}{1998}).

\bibitem[{\citenamefont{Sambandamurthy
  et~al.}(2004)\citenamefont{Sambandamurthy, Engel, Johansson, and
  Shahar}}]{Sambandamurthy:PRL2004}
\bibinfo{author}{\bibfnamefont{G.}~\bibnamefont{Sambandamurthy}},
  \bibinfo{author}{\bibfnamefont{L.~W.} \bibnamefont{Engel}},
  \bibinfo{author}{\bibfnamefont{A.}~\bibnamefont{Johansson}},
  \bibnamefont{and} \bibinfo{author}{\bibfnamefont{D.}~\bibnamefont{Shahar}},
  \bibinfo{journal}{Phys. Rev. Lett.} \textbf{\bibinfo{volume}{92}},
  \bibinfo{pages}{107005} (\bibinfo{year}{2004}).

\bibitem[{\citenamefont{Steiner et~al.}(2005)\citenamefont{Steiner, Boebinger,
  and Kapitulnik}}]{Steiner:PRL2005}
\bibinfo{author}{\bibfnamefont{M.~A.} \bibnamefont{Steiner}},
  \bibinfo{author}{\bibfnamefont{G.}~\bibnamefont{Boebinger}},
  \bibnamefont{and}
  \bibinfo{author}{\bibfnamefont{A.}~\bibnamefont{Kapitulnik}},
  \bibinfo{journal}{Phys. Rev. Lett.} \textbf{\bibinfo{volume}{94}},
  \bibinfo{pages}{107008} (\bibinfo{year}{2005}).

\bibitem[{\citenamefont{Baturina et~al.}(2007)\citenamefont{Baturina, Mironov,
  Vinokur, Baklanov, and Strunk}}]{Baturina:PRL2007}
\bibinfo{author}{\bibfnamefont{T.~I.} \bibnamefont{Baturina}},
  \bibinfo{author}{\bibfnamefont{A.~Y.} \bibnamefont{Mironov}},
  \bibinfo{author}{\bibfnamefont{V.~M.} \bibnamefont{Vinokur}},
  \bibinfo{author}{\bibfnamefont{M.~R.} \bibnamefont{Baklanov}},
  \bibnamefont{and} \bibinfo{author}{\bibfnamefont{C.}~\bibnamefont{Strunk}},
  \bibinfo{journal}{Phys. Rev. Lett.} \textbf{\bibinfo{volume}{99}},
  \bibinfo{pages}{257003} (\bibinfo{year}{2007}).

\bibitem[{\citenamefont{Sac{\'e}p{\'e}
  et~al.}(2011)\citenamefont{Sac{\'e}p{\'e}, Dubouchet, Chapelier, Sanquer,
  Ovadia, Shahar, Feigel'man, and Ioffe}}]{Sacepe:NatPhys2011}
\bibinfo{author}{\bibfnamefont{B.}~\bibnamefont{Sac{\'e}p{\'e}}},
  \bibinfo{author}{\bibfnamefont{T.}~\bibnamefont{Dubouchet}},
  \bibinfo{author}{\bibfnamefont{C.}~\bibnamefont{Chapelier}},
  \bibinfo{author}{\bibfnamefont{M.}~\bibnamefont{Sanquer}},
  \bibinfo{author}{\bibfnamefont{M.}~\bibnamefont{Ovadia}},
  \bibinfo{author}{\bibfnamefont{D.}~\bibnamefont{Shahar}},
  \bibinfo{author}{\bibfnamefont{M.}~\bibnamefont{Feigel'man}},
  \bibnamefont{and} \bibinfo{author}{\bibfnamefont{L.}~\bibnamefont{Ioffe}},
  \bibinfo{journal}{Nat. Phys.} \textbf{\bibinfo{volume}{7}},
  \bibinfo{pages}{239} (\bibinfo{year}{2011}).

\bibitem[{\citenamefont{{Stewart Jr.} et~al.}(2007)\citenamefont{{Stewart Jr.},
  Yin, Xu, and {Valles Jr.}}}]{Stewart:Science2007}
\bibinfo{author}{\bibfnamefont{M.~D.} \bibnamefont{{Stewart Jr.}}},
  \bibinfo{author}{\bibfnamefont{A.}~\bibnamefont{Yin}},
  \bibinfo{author}{\bibfnamefont{J.~M.} \bibnamefont{Xu}}, \bibnamefont{and}
  \bibinfo{author}{\bibfnamefont{J.~M.} \bibnamefont{{Valles Jr.}}},
  \bibinfo{journal}{Science} \textbf{\bibinfo{volume}{318}},
  \bibinfo{pages}{1273} (\bibinfo{year}{2007}).

\bibitem[{\citenamefont{Chand et~al.}(2012)\citenamefont{Chand, Saraswat,
  Kamlapure, Mondal, Kumar, Jesudasan, Bagwe, Benfatto, Tripathi, and
  Raychaudhuri}}]{Mondal:PRB2012}
\bibinfo{author}{\bibfnamefont{M.}~\bibnamefont{Chand}},
  \bibinfo{author}{\bibfnamefont{G.}~\bibnamefont{Saraswat}},
  \bibinfo{author}{\bibfnamefont{A.}~\bibnamefont{Kamlapure}},
  \bibinfo{author}{\bibfnamefont{M.}~\bibnamefont{Mondal}},
  \bibinfo{author}{\bibfnamefont{S.}~\bibnamefont{Kumar}},
  \bibinfo{author}{\bibfnamefont{J.}~\bibnamefont{Jesudasan}},
  \bibinfo{author}{\bibfnamefont{V.}~\bibnamefont{Bagwe}},
  \bibinfo{author}{\bibfnamefont{L.}~\bibnamefont{Benfatto}},
  \bibinfo{author}{\bibfnamefont{V.}~\bibnamefont{Tripathi}}, \bibnamefont{and}
  \bibinfo{author}{\bibfnamefont{P.}~\bibnamefont{Raychaudhuri}},
  \bibinfo{journal}{Phys. Rev. B} \textbf{\bibinfo{volume}{85}},
  \bibinfo{pages}{014508} (\bibinfo{year}{2012}).

\bibitem[{\citenamefont{Delsing et~al.}(1994)\citenamefont{Delsing, Chen,
  Haviland, Harada, and Claeson}}]{Delsing:PRB1994}
\bibinfo{author}{\bibfnamefont{P.}~\bibnamefont{Delsing}},
  \bibinfo{author}{\bibfnamefont{C.~D.} \bibnamefont{Chen}},
  \bibinfo{author}{\bibfnamefont{D.~B.} \bibnamefont{Haviland}},
  \bibinfo{author}{\bibfnamefont{Y.}~\bibnamefont{Harada}}, \bibnamefont{and}
  \bibinfo{author}{\bibfnamefont{T.}~\bibnamefont{Claeson}},
  \bibinfo{journal}{Phys. Rev. B} \textbf{\bibinfo{volume}{50}},
  \bibinfo{pages}{3959} (\bibinfo{year}{1994}).

\bibitem[{\citenamefont{Fistul et~al.}(2008)\citenamefont{Fistul, Vinokur, and
  Baturina}}]{Fistul:PRL2008}
\bibinfo{author}{\bibfnamefont{M.~V.} \bibnamefont{Fistul}},
  \bibinfo{author}{\bibfnamefont{V.~M.} \bibnamefont{Vinokur}},
  \bibnamefont{and} \bibinfo{author}{\bibfnamefont{T.~I.}
  \bibnamefont{Baturina}}, \bibinfo{journal}{Phys. Rev. Lett.}
  \textbf{\bibinfo{volume}{100}}, \bibinfo{pages}{086805}
  (\bibinfo{year}{2008}).

\bibitem[{\citenamefont{Syzranov et~al.}(2009)\citenamefont{Syzranov, Efetov,
  and Altshuler}}]{Syzranov:PRL2009}
\bibinfo{author}{\bibfnamefont{S.~V.} \bibnamefont{Syzranov}},
  \bibinfo{author}{\bibfnamefont{K.~B.} \bibnamefont{Efetov}},
  \bibnamefont{and} \bibinfo{author}{\bibfnamefont{B.~L.}
  \bibnamefont{Altshuler}}, \bibinfo{journal}{Phys. Rev. Lett.}
  \textbf{\bibinfo{volume}{103}}, \bibinfo{pages}{127001}
  (\bibinfo{year}{2009}).

\bibitem[{\citenamefont{Nguyen et~al.}(2009)\citenamefont{Nguyen, Hollen,
  Stewart, Shainline, Yin, Xu, and Valles}}]{Nguyen:PRL2009}
\bibinfo{author}{\bibfnamefont{H.~Q.} \bibnamefont{Nguyen}},
  \bibinfo{author}{\bibfnamefont{S.~M.} \bibnamefont{Hollen}},
  \bibinfo{author}{\bibfnamefont{M.~D.} \bibnamefont{Stewart}},
  \bibinfo{author}{\bibfnamefont{J.}~\bibnamefont{Shainline}},
  \bibinfo{author}{\bibfnamefont{A.}~\bibnamefont{Yin}},
  \bibinfo{author}{\bibfnamefont{J.~M.} \bibnamefont{Xu}}, \bibnamefont{and}
  \bibinfo{author}{\bibfnamefont{J.~M.} \bibnamefont{Valles}},
  \bibinfo{journal}{Phys. Rev. Lett.} \textbf{\bibinfo{volume}{103}},
  \bibinfo{pages}{157001} (\bibinfo{year}{2009}).

\bibitem[{\citenamefont{Welp et~al.}(2002)\citenamefont{Welp, Xiao, Jiang,
  Vlasko-Vlasov, Bader, Crabtree, Liang, Chik, and Xu}}]{Welp:PRB2002}
\bibinfo{author}{\bibfnamefont{U.}~\bibnamefont{Welp}},
  \bibinfo{author}{\bibfnamefont{Z.~L.} \bibnamefont{Xiao}},
  \bibinfo{author}{\bibfnamefont{J.~S.} \bibnamefont{Jiang}},
  \bibinfo{author}{\bibfnamefont{V.~K.} \bibnamefont{Vlasko-Vlasov}},
  \bibinfo{author}{\bibfnamefont{S.~D.} \bibnamefont{Bader}},
  \bibinfo{author}{\bibfnamefont{G.~W.} \bibnamefont{Crabtree}},
  \bibinfo{author}{\bibfnamefont{J.}~\bibnamefont{Liang}},
  \bibinfo{author}{\bibfnamefont{H.}~\bibnamefont{Chik}}, \bibnamefont{and}
  \bibinfo{author}{\bibfnamefont{J.~M.} \bibnamefont{Xu}},
  \bibinfo{journal}{Phys. Rev. B} \textbf{\bibinfo{volume}{66}},
  \bibinfo{pages}{212507} (\bibinfo{year}{2002}).

\bibitem[{\citenamefont{Dubi et~al.}(2007)\citenamefont{Dubi, Meir, and
  Avishai}}]{Dubi:Nature2007}
\bibinfo{author}{\bibfnamefont{Y.}~\bibnamefont{Dubi}},
  \bibinfo{author}{\bibfnamefont{Y.}~\bibnamefont{Meir}}, \bibnamefont{and}
  \bibinfo{author}{\bibfnamefont{Y.}~\bibnamefont{Avishai}},
  \bibinfo{journal}{Nature} \textbf{\bibinfo{volume}{449}},
  \bibinfo{pages}{876} (\bibinfo{year}{2007}).

\bibitem[{\citenamefont{Feigel'man et~al.}(2010)\citenamefont{Feigel'man,
  Ioffe, Kravtsov, and Cuevas}}]{Feigelman:APhys2010}
\bibinfo{author}{\bibfnamefont{M.~V.} \bibnamefont{Feigel'man}},
  \bibinfo{author}{\bibfnamefont{L.~B.} \bibnamefont{Ioffe}},
  \bibinfo{author}{\bibfnamefont{V.~E.} \bibnamefont{Kravtsov}},
  \bibnamefont{and} \bibinfo{author}{\bibfnamefont{E.}~\bibnamefont{Cuevas}},
  \bibinfo{journal}{Ann. Phys.} \textbf{\bibinfo{volume}{325}},
  \bibinfo{pages}{1390} (\bibinfo{year}{2010}).

\bibitem[{\citenamefont{Pokrovsky et~al.}(2010)\citenamefont{Pokrovsky, Falco,
  and Nattermann}}]{Falco:PRL2010}
\bibinfo{author}{\bibfnamefont{V.~L.} \bibnamefont{Pokrovsky}},
  \bibinfo{author}{\bibfnamefont{G.~M.} \bibnamefont{Falco}}, \bibnamefont{and}
  \bibinfo{author}{\bibfnamefont{T.}~\bibnamefont{Nattermann}},
  \bibinfo{journal}{Phys. Rev. Lett.} \textbf{\bibinfo{volume}{105}},
  \bibinfo{pages}{267001} (\bibinfo{year}{2010}).

\bibitem[{\citenamefont{Bouadim et~al.}(2011)\citenamefont{Bouadim, Loh,
  Randeria, and Trivedi}}]{Bouadim:NatPhys2011}
\bibinfo{author}{\bibfnamefont{K.}~\bibnamefont{Bouadim}},
  \bibinfo{author}{\bibfnamefont{Y.~L.} \bibnamefont{Loh}},
  \bibinfo{author}{\bibfnamefont{M.}~\bibnamefont{Randeria}}, \bibnamefont{and}
  \bibinfo{author}{\bibfnamefont{N.}~\bibnamefont{Trivedi}},
  \bibinfo{journal}{Nat. Phys.} \textbf{\bibinfo{volume}{7}},
  \bibinfo{pages}{1} (\bibinfo{year}{2011}).

\bibitem[{\citenamefont{Gantmakher and Dolgopolov}(2010)}]{Gantmakher:PUsp2010}
\bibinfo{author}{\bibfnamefont{V.~F.} \bibnamefont{Gantmakher}}
  \bibnamefont{and} \bibinfo{author}{\bibfnamefont{V.~T.}
  \bibnamefont{Dolgopolov}}, \bibinfo{journal}{Phys.-Usp.}
  \textbf{\bibinfo{volume}{53}}, \bibinfo{pages}{1} (\bibinfo{year}{2010}).

\bibitem[{\citenamefont{Ghosal et~al.}(1998)\citenamefont{Ghosal, Randeria, and
  Trivedi}}]{Ghosal:PRL1998}
\bibinfo{author}{\bibfnamefont{A.}~\bibnamefont{Ghosal}},
  \bibinfo{author}{\bibfnamefont{M.}~\bibnamefont{Randeria}}, \bibnamefont{and}
  \bibinfo{author}{\bibfnamefont{N.}~\bibnamefont{Trivedi}},
  \bibinfo{journal}{Phys. Rev. Lett.} \textbf{\bibinfo{volume}{81}},
  \bibinfo{pages}{3940} (\bibinfo{year}{1998}).

\bibitem[{\citenamefont{Hollen et~al.}(2011)\citenamefont{Hollen, Nguyen,
  Rudisaile, Stewart, Shainline, Xu, and {Valles Jr.}}}]{Hollen:PRB2011}
\bibinfo{author}{\bibfnamefont{S.~M.} \bibnamefont{Hollen}},
  \bibinfo{author}{\bibfnamefont{H.~Q.} \bibnamefont{Nguyen}},
  \bibinfo{author}{\bibfnamefont{E.}~\bibnamefont{Rudisaile}},
  \bibinfo{author}{\bibfnamefont{M.~D.} \bibnamefont{Stewart}},
  \bibinfo{author}{\bibfnamefont{J.}~\bibnamefont{Shainline}},
  \bibinfo{author}{\bibfnamefont{J.~M.} \bibnamefont{Xu}}, \bibnamefont{and}
  \bibinfo{author}{\bibfnamefont{J.~M.} \bibnamefont{{Valles Jr.}}},
  \bibinfo{journal}{Phys. Rev. B} \textbf{\bibinfo{volume}{84}},
  \bibinfo{pages}{064528} (\bibinfo{year}{2011}).

\bibitem[{\citenamefont{Hollen and {Valles Jr.}}(2012)}]{Hollen:JPhys2012}
\bibinfo{author}{\bibfnamefont{S.~M.} \bibnamefont{Hollen}} \bibnamefont{and}
  \bibinfo{author}{\bibfnamefont{J.~M.} \bibnamefont{{Valles Jr.}}},
  \bibinfo{journal}{Journal of Physics: Conference Series}
  \textbf{\bibinfo{volume}{376}}, \bibinfo{pages}{012002}
  (\bibinfo{year}{2012}).

\bibitem[{\citenamefont{Ekinci and {Valles Jr.}}(1998)}]{Ekinci:PRB1998}
\bibinfo{author}{\bibfnamefont{K.~L.} \bibnamefont{Ekinci}} \bibnamefont{and}
  \bibinfo{author}{\bibfnamefont{J.~M.} \bibnamefont{{Valles Jr.}}},
  \bibinfo{journal}{Phys. Rev. B} \textbf{\bibinfo{volume}{58}},
  \bibinfo{pages}{7347} (\bibinfo{year}{1998}).

\bibitem[{\citenamefont{Liu et~al.}(1993)\citenamefont{Liu, Haviland, Nease,
  and Goldman}}]{Liu:PRB1993}
\bibinfo{author}{\bibfnamefont{Y.}~\bibnamefont{Liu}},
  \bibinfo{author}{\bibfnamefont{D.~B.} \bibnamefont{Haviland}},
  \bibinfo{author}{\bibfnamefont{B.}~\bibnamefont{Nease}}, \bibnamefont{and}
  \bibinfo{author}{\bibfnamefont{A.~M.} \bibnamefont{Goldman}},
  \bibinfo{journal}{Phys. Rev. B} \textbf{\bibinfo{volume}{47}},
  \bibinfo{pages}{5931} (\bibinfo{year}{1993}).

\bibitem[{\citenamefont{Dynes et~al.}(1978)\citenamefont{Dynes, Garno, and
  Rowell}}]{Dynes:PRL1978}
\bibinfo{author}{\bibfnamefont{R.~C.} \bibnamefont{Dynes}},
  \bibinfo{author}{\bibfnamefont{J.~P.} \bibnamefont{Garno}}, \bibnamefont{and}
  \bibinfo{author}{\bibfnamefont{J.~M.} \bibnamefont{Rowell}},
  \bibinfo{journal}{Phys. Rev. Lett.} \textbf{\bibinfo{volume}{40}},
  \bibinfo{pages}{479} (\bibinfo{year}{1978}).

\end{thebibliography}
\end{document}